\documentclass{PoS}
\usepackage{bm} % bold math
\usepackage{amsfonts} % AMS
\usepackage{amssymb} % AMS
\usepackage{amsmath} % AMS
\usepackage{graphicx}    % needed for including graphics e.g. EPS, PS
\usepackage{subfigure}
\usepackage[export]{adjustbox}
\usepackage{multirow}
\usepackage{mathtools}

\newcommand{\MSb}{{\overline{MS}}}

\title{Nucleon charges and form factors using clover and HISQ ensembles}

\ShortTitle{Nucleon charges and form factors using clover and HISQ ensembles}

\author{\speaker{Sungwoo Park}, Tanmoy Bhattacharya, Rajan Gupta \\
        T-2, Los Alamos National Laboratory, Los Alamos, NM 87545, USA\\
        E-mail: \email{sungwoo@lanl.gov}, \email{tanmoy@lanl.gov}, \email{rg@lanl.gov}}

\author{Yong-Chull Jang
  \\ Phsics Department, Brookhaven National Laboratory, Upton, NY 11973, USA
  \\ E-mail: \email{ypj@bnl.gov}}

\author{Balint Joo 
 \\ Thomas Jefferson National Accelerator Facility, Newport News, VA 23606, USA
 \\ E-mail: \email{bjoo@jlab.org}}

\author{Huey-Wen Lin
  \\ Department of Physics and Astronomy, Michigan State University, East Lansing, MI 48824, USA
  \\ E-mail: \email{hwlin@pa.msu.edu}}

\author{Boram Yoon\\
        CCS-7, Los Alamos National Laboratory, Los Alamos, NM 87545, USA\\
        E-mail: \email{boram@lanl.gov}}

\abstract{We present high statistics ($\mathcal{O}(2\times 10^5)$
  measurements) preliminary results on (i) the isovector charges,
  $g^{u-d}_{A,S,T}$, and form factors, $G^{u-d}_E(Q^2)$,
  $G^{u-d}_M(Q^2)$, $G^{u-d}_A(Q^2)$, $\widetilde G^{u-d}_P(Q^2)$,
  $G^{u-d}_P(Q^2)$, on six $2+1$-flavor Wilson-clover ensembles
  generated by the JLab/W\&M/LANL/MIT collaboration with lattice
  parameters given in Table~\ref{tab:ens}. Examples of the impact of
  using different estimates of the excited state spectra are given for
  the clover-on-clover data, and as discussed in~\cite{Jang:2019vkm},
  the biggest difference on including the lower energy (close to
  $N\pi$ and $N\pi\pi$) states is in the axial channel.  (ii) Flavor
  diagonal axial, tensor and scalar charges, $g^{u,d,s}_{A,S,T}$, 
  are calculated with the clover-on-HISQ formulation using nine
  2+1+1-flavor HISQ ensembles generated by the MILC
  collaboration~\cite{Bazavov:2012xda} with lattice parameters given
  in Table~\ref{tab:hisq}. Once finished, the calculations of
  $g^{u,d,s}_{A,T}$ will update the results given in
  Refs.~\cite{Lin:2018obj,Gupta:2018lvp}. The estimates for
  $g^{u,d,s}_{S}$ and $\sigma_{N\pi}$ are new. Overall, a large part of the
  focus is on understanding the excited state contamination (ESC), and
  the results discussed provide a partial status report on developing
  defensible analyses strategies that include contributions of
  possible low-lying excited states to individual nucleon matrix
  elements.  }

\FullConference{37th International Symposium on Lattice Field Theory - Lattice2019\\
		16-22 June 2019\\
		Wuhan, China}

\begin{document}
%%============================================================================
%%============================================================================
\section{Isovector charges with 2+1-flavor clover fermions}
%%============================================================================
%%============================================================================

\begin{figure}[t]  %1
  \vspace{-5mm}
  \includegraphics[width=0.325\textwidth]{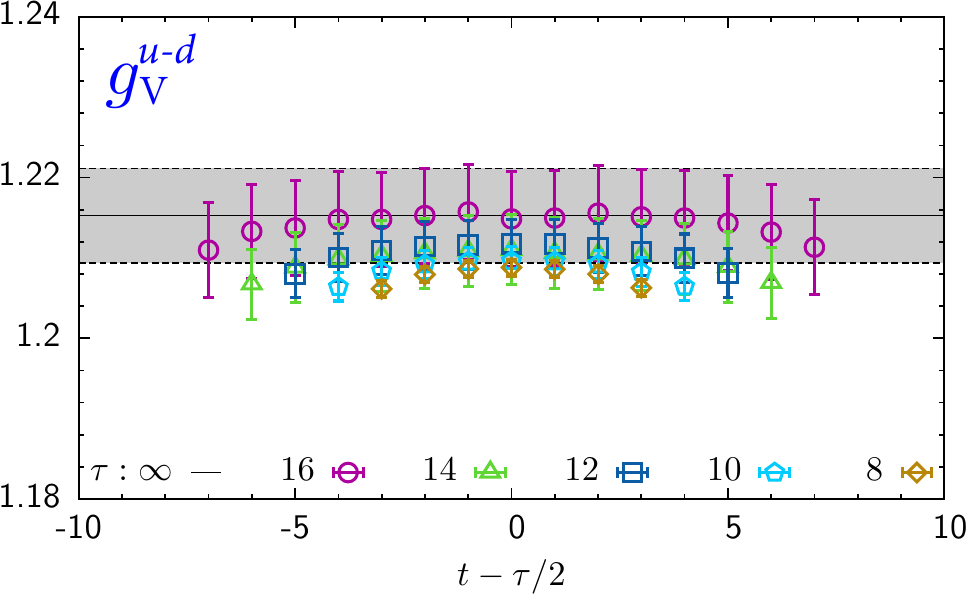}
  \includegraphics[width=0.325\textwidth]{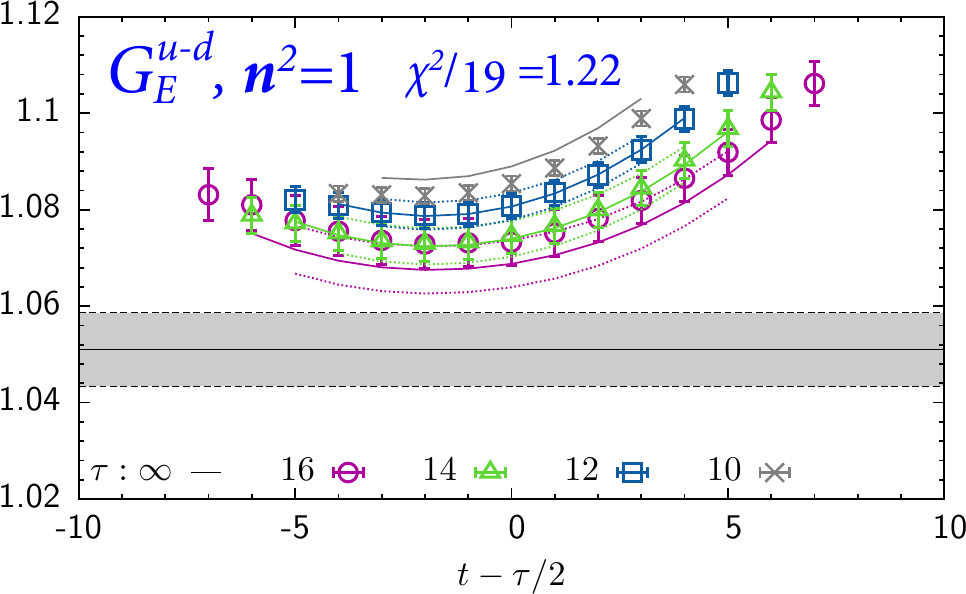}
  \includegraphics[width=0.325\textwidth]{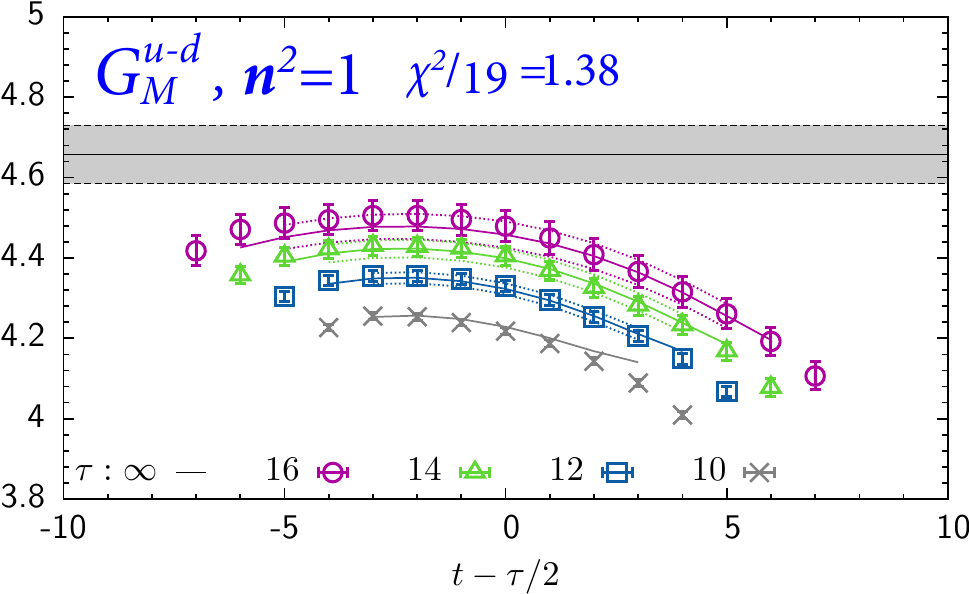} \\
  \vspace{-5mm}
  \caption{Example of ESC in unrenormalized isovector vector
    charge $g_{V}^{u-d}$ and form factors $G_{E,M}^{u-d}(\vec
    p^2)$ at $\vec{p}^2=(2\pi/L)^2\vec n^2$ with $\vec n^2=1$ on $a09m170L$
    clover lattices. For $g_V^{u-d}$, the $\tau\to\infty$ value (grey band) is 
    the average of the 5 middle data points with $\tau=16$. Fits to 
    $G_{E,M}^{u-d}$ use the $\{4,3^\ast\}$ strategy.  }
  \label{figs:V}
\end{figure}

Examples of ESC in the vector charge, $g_V^{u-d}$, and form factors $G_E$ and $G_M$
are illustrated in
Fig.~\ref{figs:V}. $g_V^{u-d}$ does not vary monotonically with source-sink
separation $\tau$, however it is constant to within 1--2\%.  So we take the
average of the central points with the largest $\tau$ (the plateau method).  With this choice
the identity $Z_Vg_V^{u-d}=1$ is satisfied to within $ 3\%$ with
the $Z_V$ calculated in Ref.~\cite{Yoon:2016jzj}. 

\begin{figure}[t] %2
  \vspace{-5mm}
  \center
  \includegraphics[width=0.325\textwidth]{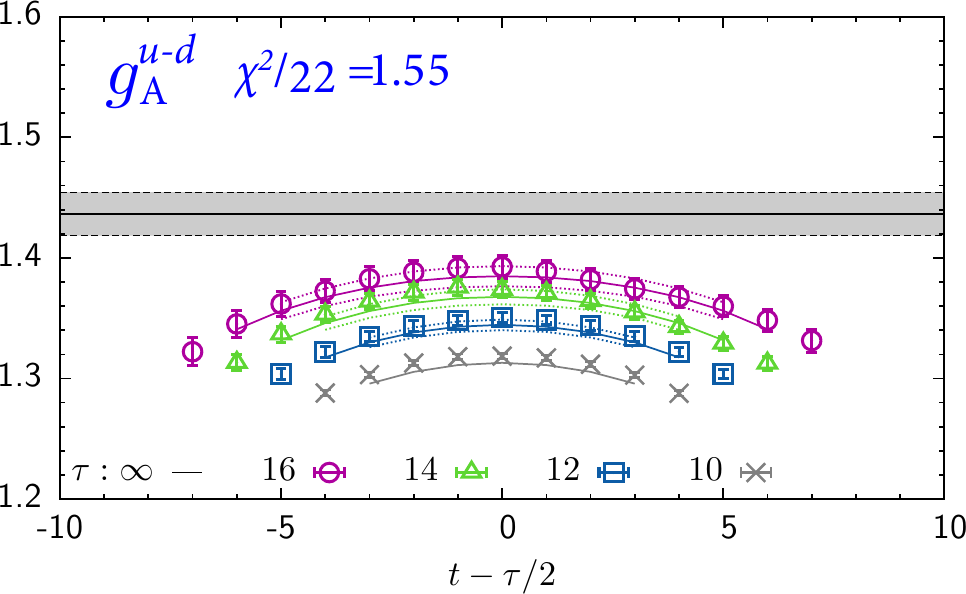}
  \includegraphics[width=0.325\textwidth]{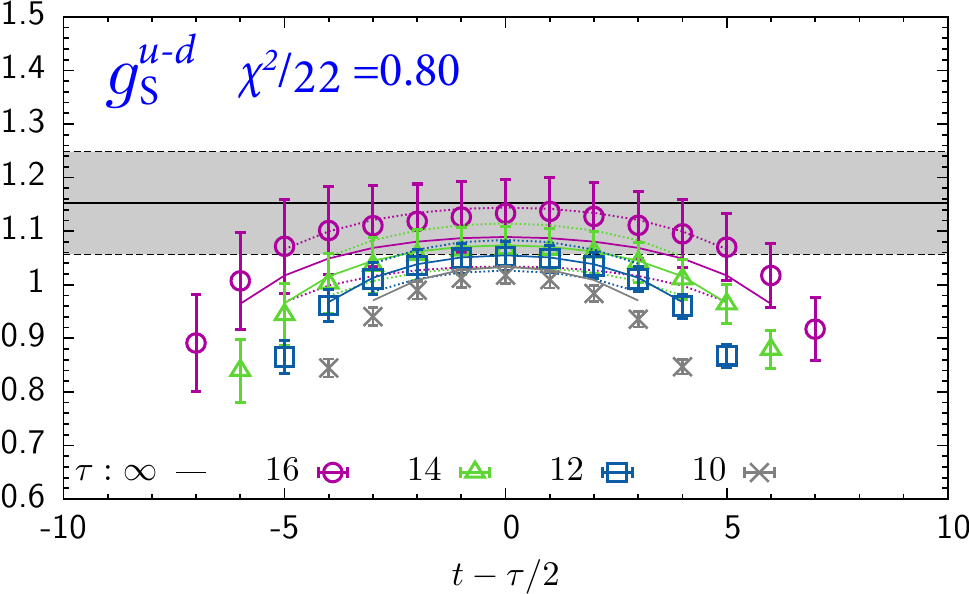}
  \includegraphics[width=0.325\textwidth]{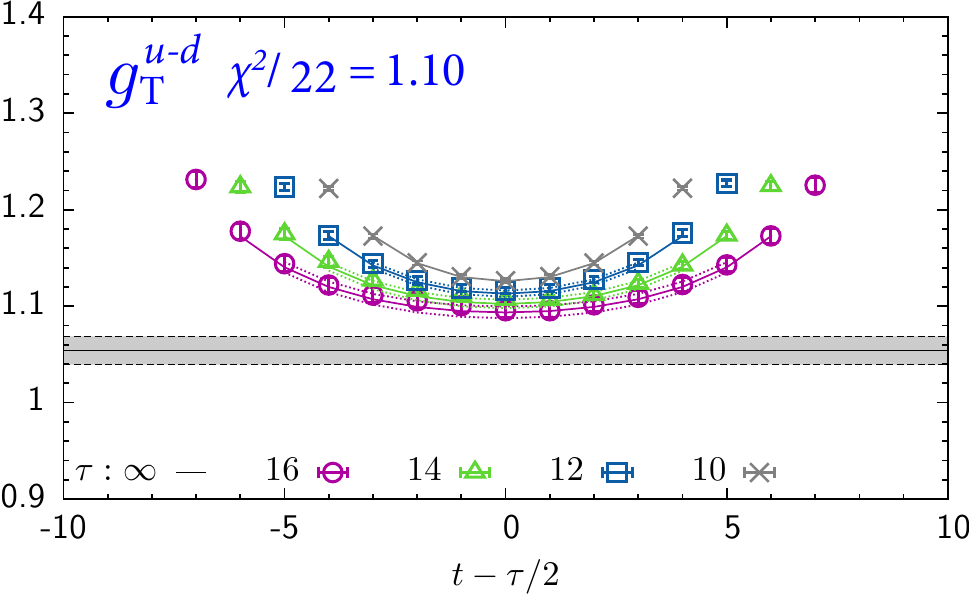} 
\\
  \includegraphics[width=0.325\textwidth]{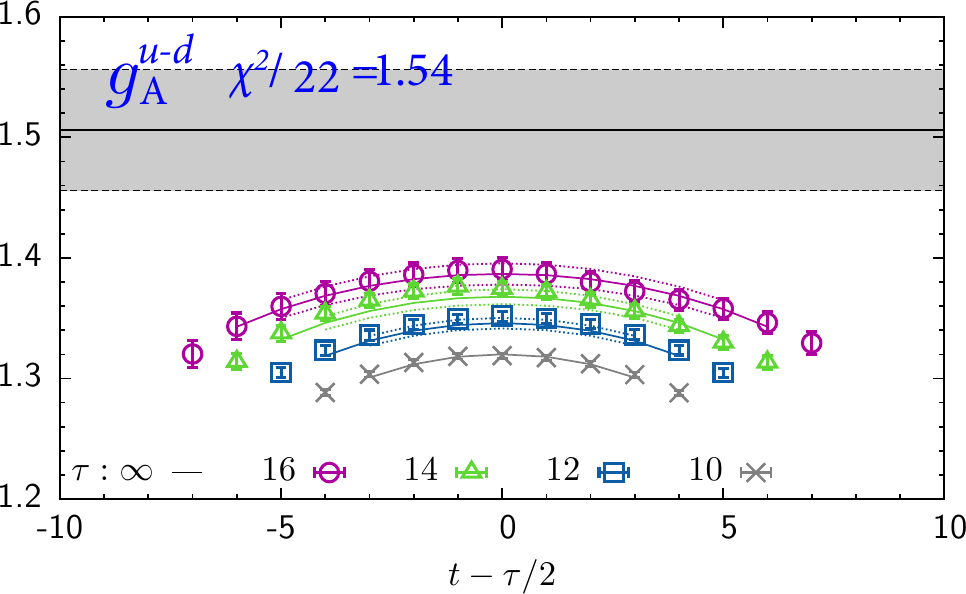}
  \includegraphics[width=0.325\textwidth]{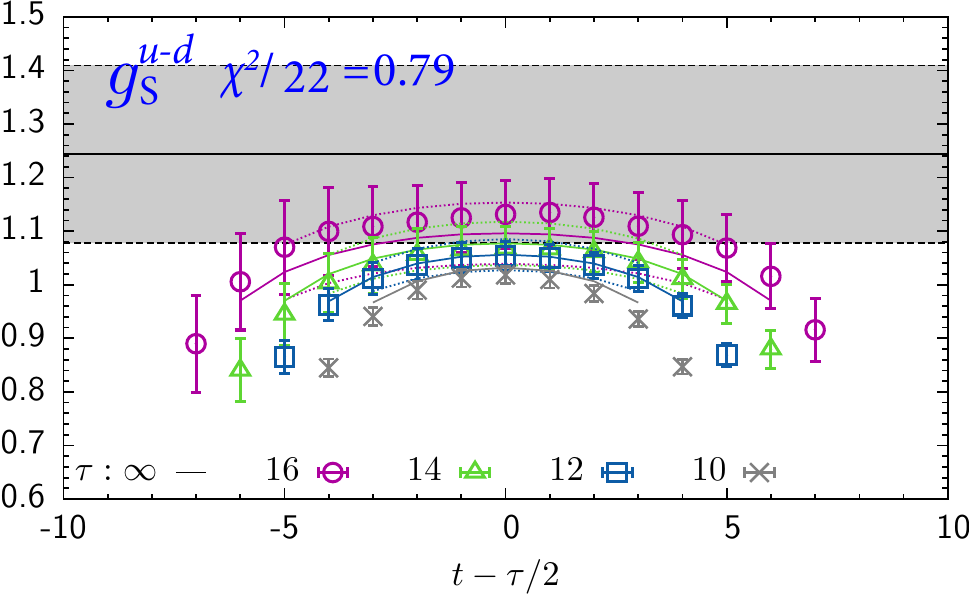}
  \includegraphics[width=0.325\textwidth]{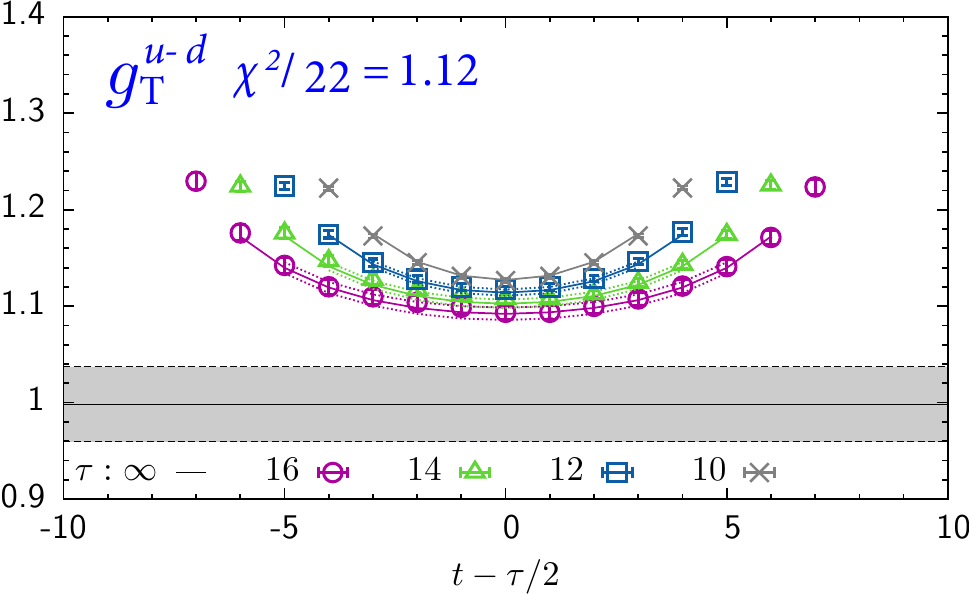}
  \vspace{-2mm}
  \caption{Data and ESC fits for unrenormalized charges
    $g_{A,S,T}^{u-d}$ on $a09m170L$ clover lattices using the 
    $\{4,3^\ast\}$ fit (top 3 panels) and the $\{3^{N \pi},3^\ast\}$ fit (bottom 3 panels). 
    Values of $\tau$ and [$\chi^2/$DOF] are given in the legend.}
  \label{figs:clov-charges}
\end{figure}

Data for the charges, $g_{A,S,T}^{u-d}$, show significant ESC as
discussed in~\cite{Yoon:2016jzj,Gupta:2018qil}.  As described in
Ref.~\cite{Jang:2019vkm}, the key parameter controlling ESC is the
energy, $E_1$, of the first excited state. Its value, obtained from a
4-state fit to the 2-point function using emperical Bayesian
priors~\cite{Rajan:2017lxk}, is much larger than that of
non-interating $N\pi$ or $N\pi\pi$-states, especially for physical
$M_\pi$ ensembles. Using the energy of the $N\pi$ state as a prior for
$E_1$ in a 3-state fit gives a much lower output value for $E_1$ but
with an equally good $\chi^2/$DOF, indicating a flat direction in the
parameter space. Note that with a small $E_1$, even $E_0$ is slightly
smaller. We have, therefore, analyzed the ESC using multiple
strategies and, here, compare two for $g_{A,S,T}^{u-d}$ based on $3^\ast$-state fits
($3$-state truncation of the spectral decomposition of the 3-point
functions with $\langle 2'| \mathcal{O} | 2\rangle =0$). The standard
$\{4,3^\ast\}$ and $\{3^{N\pi},3^\ast\}$. In $\{4,3^\ast\}$, the
spectrum is taken from the standard 4-state fit~\cite{Jang:2019jkn}. In
$\{3^{N\pi},3^\ast\}$, the energy $E^{N\pi}_1$ of the lowest possible
state, $N(\mathbf{1})\pi(-\mathbf{1})$, is used as a prior for $E_1$ in a 3-state
fit and the resulting outputs, ground-state amplitude $A_0$ and energies 
$E_0$, $E_1$ and $E_2$, are used as inputs in fits to the 3-point
functions. The data and fits for $\{4,3^\ast\}$ and
$\{3^{N\pi},3^\ast\}$ are compared in Fig.~\ref{figs:clov-charges} for
the $a091m170L$ ensemble, where one expects the largest effect as it
has the smallest $M_\pi \sim 170$~MeV and $Q_{\rm min}^2$, with
$Q^2=\vec p^2 - (E_N-M_N)^2$ being the Euclidean 4-momentum squared transferred.

\begin{figure}[t]  %3
%  \vspace{-5mm}
  \includegraphics[width=0.325\textwidth]{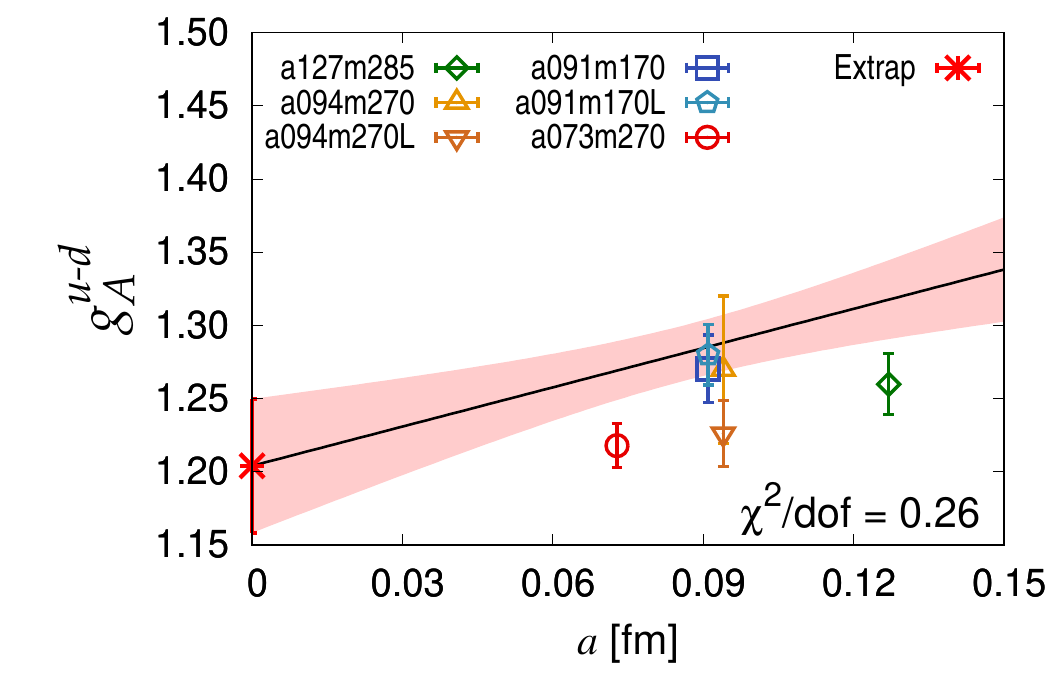}
  \includegraphics[width=0.325\textwidth]{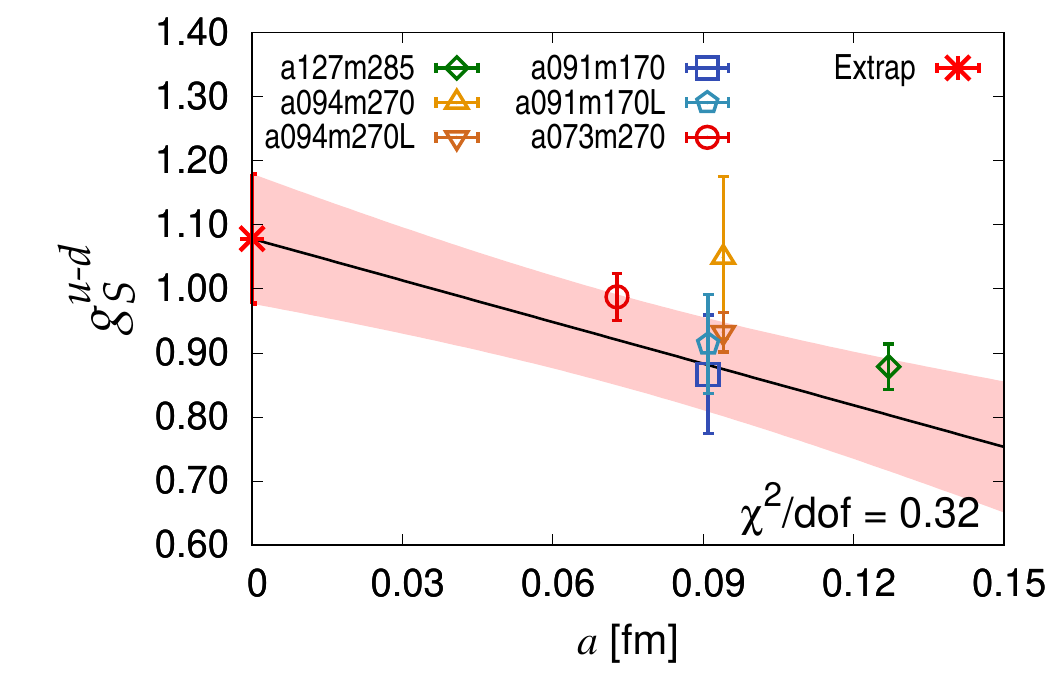}
  \includegraphics[width=0.325\textwidth]{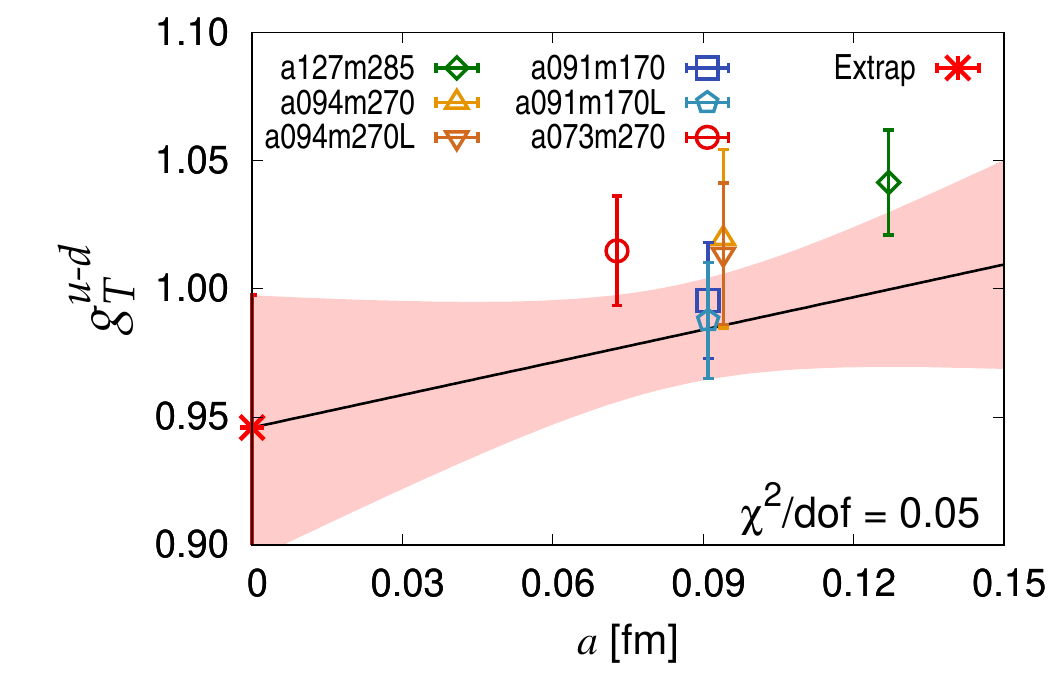}
  \includegraphics[width=0.325\textwidth]{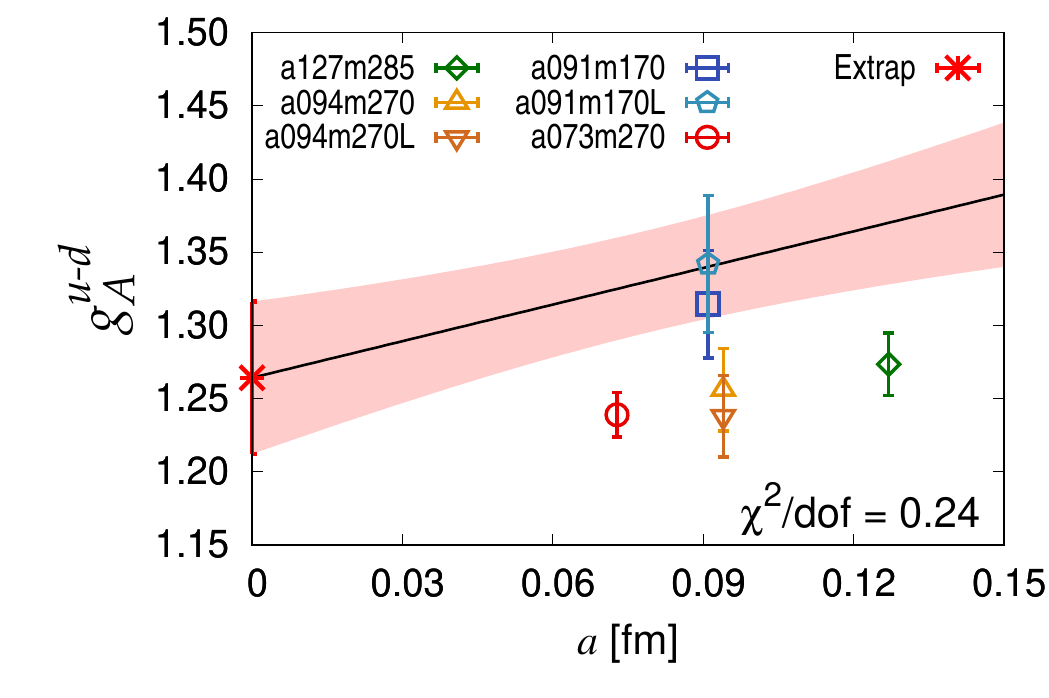}
  \includegraphics[width=0.325\textwidth]{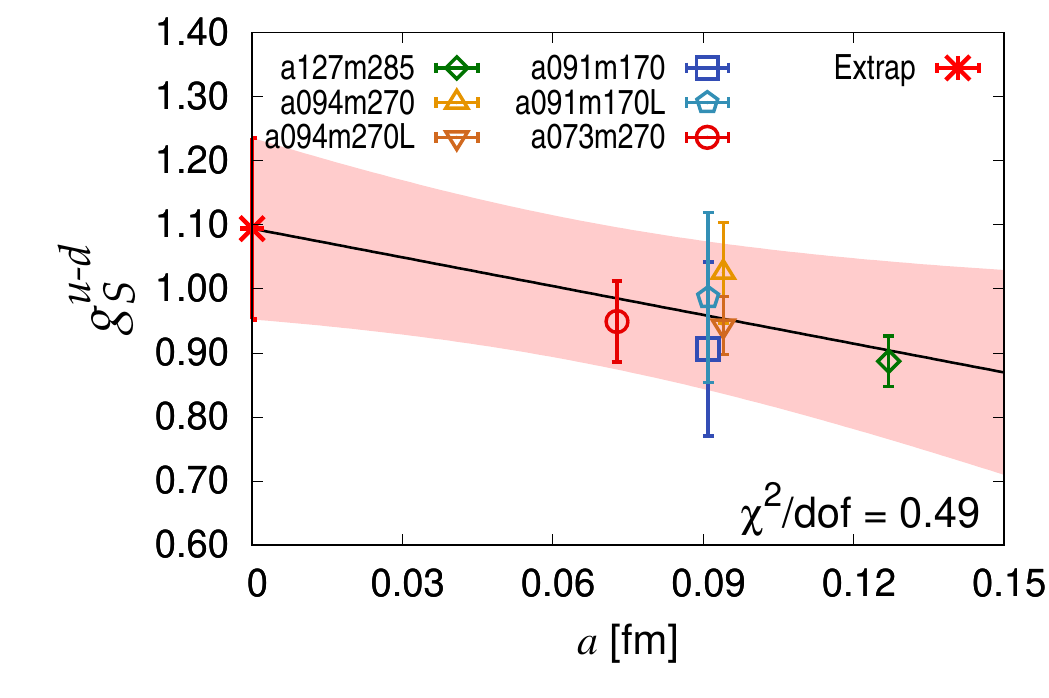}
  \includegraphics[width=0.325\textwidth]{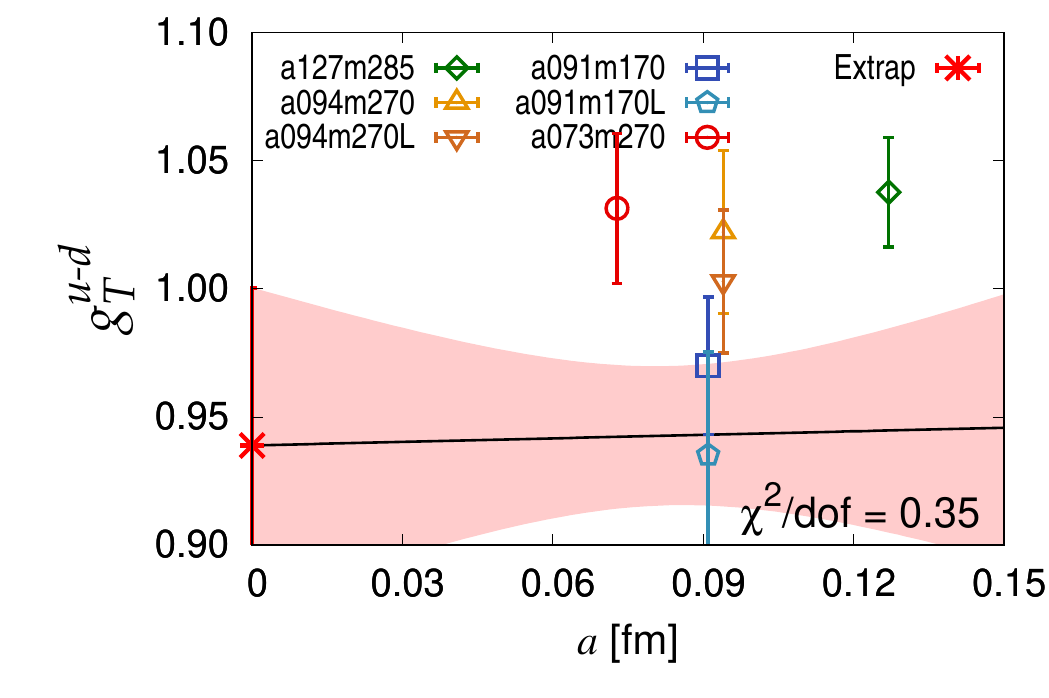}
  \vspace{-2mm}
  \caption{Chiral-continuum extrapolation of the renormalized (in
    $\MSb$ at $2$~GeV) isovector charges using the ansatz
    $f(a,M_\pi)=c_1+c_2a +c_3 M_\pi^2$.  Results with $\{4,3^\ast\}$
    ($\{3^{N\pi},3^\ast\}$) strategy are shown in the top (bottom) 3
    panels.  In each pannel, the pink band shows the result of the
    simultaneous fit plotted versus the lattice spacing $a$ with
    $M_\pi$ set to $135$~MeV. The value in the continuum limit, $a=0$,
    is marked with a red star.  }
  \label{figs:clov-charges_extrap}
\end{figure}

The value of $g_{A}^{u-d}$ is sensitive to input $E_1$ used in the ESC
fits, however, different fits are not distinguished by $\chi^2/$DOF,
again indicating a flat direction.  Renormalized charges in the $\MSb$
scheme at 2~GeV, $g_{A,S,T}^{u-d}|_R = Z_{A,S,T}^{u-d}
g_{A,S,T}^{u-d}$, are obtained using $Z_{A,S,T}^{u-d}$ from
Ref.~\cite{Yoon:2016jzj}.  Their chiral-continuum (CC) extrapolation
is done using the ansatz $f(a,M_\pi)=c_1+c_2 a + c_3 M_\pi^2$ (see
Fig.~\ref{figs:clov-charges_extrap}), and the results at
$M_\pi=135$~MeV and $a=0$ are given in Table~\ref{tab:IVcharges}.  The
difference in $g_{A}^{u-d}$ is a measure of the systematic uncertainty
associated with ESC fits. Data for $g_{S,T}^{u-d}$ from the two strategies,
shown in Fig.~\ref{figs:clov-charges_extrap} and the extrapolated
values in Table~\ref{tab:IVcharges}, are consistent within $1\sigma$
and the ESC fits do not prefer the low $E^{N\pi}_1$. \looseness-1

\begin{table}[b]
\begin{center}
%\vspace{-5mm}
\resizebox{.95\textwidth}{!}{
\begin{tabular}{l|llllrrrc}
Ensemble ID    & $a$ (fm) & $M_\pi$ (MeV) & $L^3\times T$    & $M_\pi L$  & $N_\text{conf}$ & $N^\text{HP}_\text{meas}$ & $N^\text{LP}_\text{meas}$  & $\tau$ \\\hline
$a127m285 $    & 0.127(2) & 285(3)        & $32^3\times 96$  & 5.85       & 2002   & 8008  & 256,256  & \{8, 10, 12, 14\}     \\
$a094m270 $    & 0.094(1) & 270(3)        & $32^3\times 64$  & 4.11       & 2469   & 7407  & 237,024  & \{8, 10, 12, 14, 16\}     \\
$a094m270L$    & 0.094(1) & 269(3)        & $48^3\times 128$ & 6.16       & 1854   & 7416  & 237,312  & \{8, 10, 12, 14, 16, 18\}     \\
$a091m170 $    & 0.091(1) & 170(2)        & $48^3\times 96$  & 3.7        & 2754   & 11016 & 352,512  & \{8, 10, 12, 14, 16\}     \\
$a091m170L$    & 0.091(1) & 170(2)        & $64^3\times 128$ & 5.08       & 1825   & 9125  & 292,000  & \{8, 10, 12, 14, 16\}     \\
$a073m270 $    & 0.0728(8)& 272(3)        & $48^3\times 128$ & 4.8        & 2454   & 9816  & 314,112  & \{11, 13, 15, 17, 19\} \\
\end{tabular}}
\end{center}
\vspace{-4mm}
\caption{Lattice parameters of $2+1$-flavor clover ensembles generated
  by the JLab/W\&M/LANL/MIT collaboration.  $N^\text{LP}_\text{meas}$
  low-precision and $N^\text{HP}_\text{meas}$ high-precision
  measurements of 2- and 3-point functions are made using the bias
  corrected truncated solver method (see Ref.~\cite{Yoon:2016jzj} for
  details.).  $\tau$ gives the source-sink separations
  studied. Statistics on $a094m270L$, $a091m170$ and $a091m170L$ ensembles are
  being increased. }
\label{tab:ens}
\end{table}

\begin{table}[b]
  \vspace{-2mm}
\center  
\resizebox{0.9\textwidth}{!}{
\begin{tabular}{l|cccc|ccccccccccccccccc}
Ensemble ID    & $a$ (fm) & $M_\pi$ (MeV)    & $M_\pi L$  &  $L^3\times T$ & $N_\text{conf}^l$ & $N_\text{src}^l$ & $N_\text{conf}^s$ &
    $N_\text{src}^s$ & $N_\text{LP}/N_\text{HP}$\\\hline
$a15m310$    & 0.1510(20) & 320(5)      & 3.93    & $16^3\times 48$   & 1917 & 2000 & 1919 & 2000 & 50  \\
$a12m310$    & 0.1207(11) & 310(3)      & 4.55    & $24^3\times 64$   & 1013 & 5000 & 1013 & 1500 & 30    \\
$a12m220$    & 0.1184(10) & 228(2)      & 4.38    & $32^3\times 64$   & 958  & 11000 & 958 & 4000 & 30     \\
$a09m310$    & 0.0888(8) &  313(3)      & 4.51    & $32^3\times 96$   & 1081 & 4000 & 1081 & 2000 & 30    \\
$a09m220$    & 0.0872(7) &  226(2)      & 4.79    & $48^3\times 96$   & 712  & 8000 & 847 & 10000 & 30/50  \\
$a09m130$    & 0.0871(6) &  138(1)      & 3.90    & $64^3\times 96$   & 1270 & 10000 & 877 & 10000 & 50  \\
$a06m310$    & 0.0582(4) &  320(2)      & 3.90    & $48^3\times 144$  & 830  & 4000 & 956 & 10000 & 50    \\
$a06m220$    & 0.0578(4) &  235(2)      & 4.41    & $64^3\times 144$  & 593  & 10000 & 554 & 10000 & 50\\
$a06m135$    & 0.0570(1) &  136(1)      & 3.7     & $96^3\times 192$  & 553  & 500 & 553 & 500 & 50
\end{tabular}}
\vspace{-2mm}
\caption{Parameters of the $2+1+1$-flavor HISQ ensembles used for the
  calculation of disconnected contributions (update of work in Refs.~\cite{Lin:2018obj,Gupta:2018lvp}).  $N_\text{conf}^{l,s}$
  gives the number of gauge configurations analyzed for light ($l$)
  and strange ($s$) flavors. $N_\text{src}^{l,s}$ the number of random
  sources used per configurations, and $N_\text{LP}/N_\text{HP}$ the
  ratio of low- to high-precision meausurements. Results for the connected 
  contributions are taken from Ref.~\cite{Gupta:2018qil}. }
\label{tab:hisq}
\end{table}
%%============================================================================

\begin{table}[b]
\begin{center}
%\vspace{-5mm}
\resizebox{.43\textwidth}{!}{
\begin{tabular}{l|cc}
Charge          &  $\{4,3^\ast\}$     &  $\{3^{N\pi},3^\ast\}$         \\
\hline
$g_A^{u-d}|_R$  &   1.20(5)  [0.26]   &   1.26(5)  [0.24]       \\
$g_S^{u-d}|_R$  &   1.08(10) [0.32]   &   1.09(14) [0.49]       \\
$g_T^{u-d}|_R$  &   0.95(5)  [0.05]   &   0.94(6)  [0.35]       \\
\end{tabular}
}
\end{center}
\vspace{-4mm}
\caption{$g_{A,S,T}^{u-d}|_R $ in $\overline{MS}$ scheme
  at 2~GeV calculated in 2 ways to remove ESC, and [$\chi^2/$DOF] of CC fits. }
\label{tab:IVcharges}
\end{table}

%%============================================================================
\section{Form factors}

\begin{figure}[t]
  \center
  \vspace{-5mm}
  \includegraphics[width=0.325\textwidth]{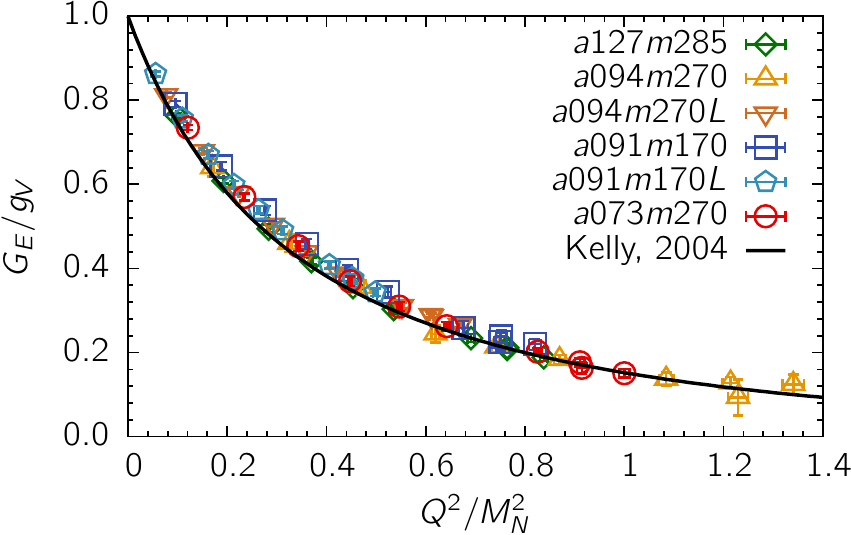}
  \includegraphics[width=0.325\textwidth]{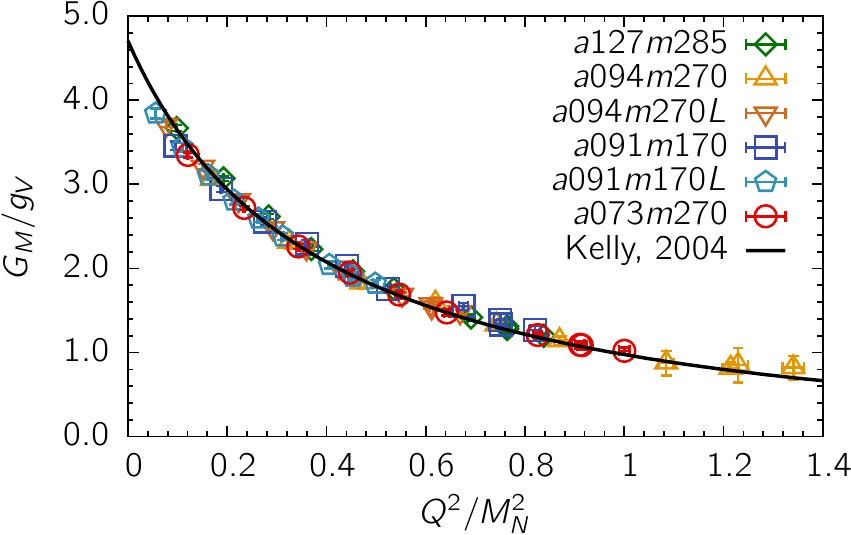}

  \includegraphics[width=0.325\textwidth]{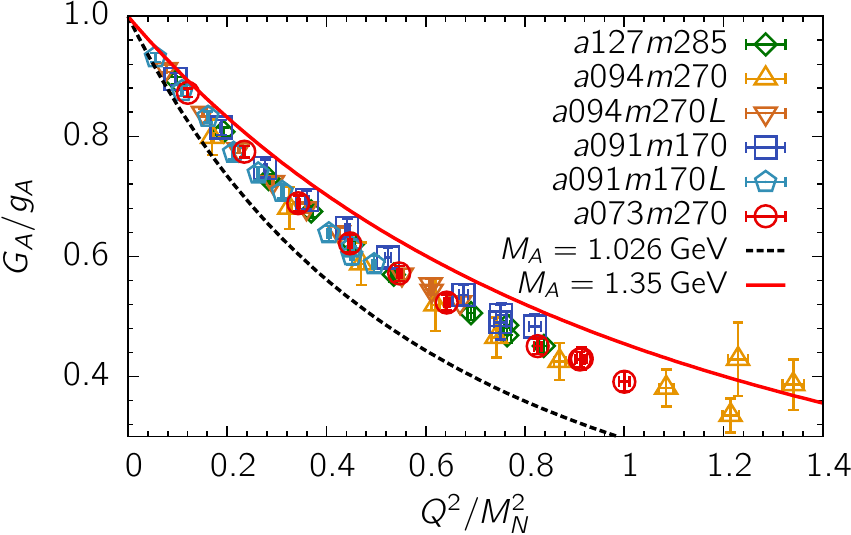}
  \includegraphics[width=0.325\textwidth]{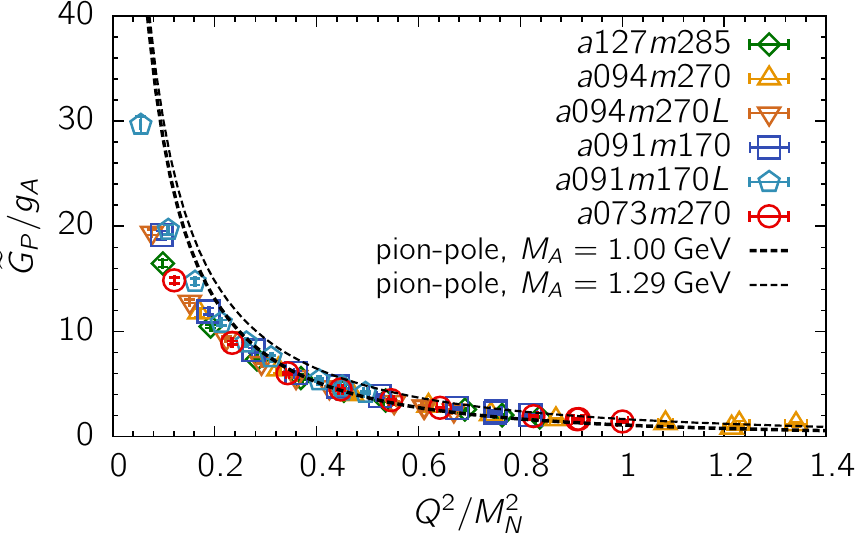}
  \includegraphics[width=0.325\textwidth]{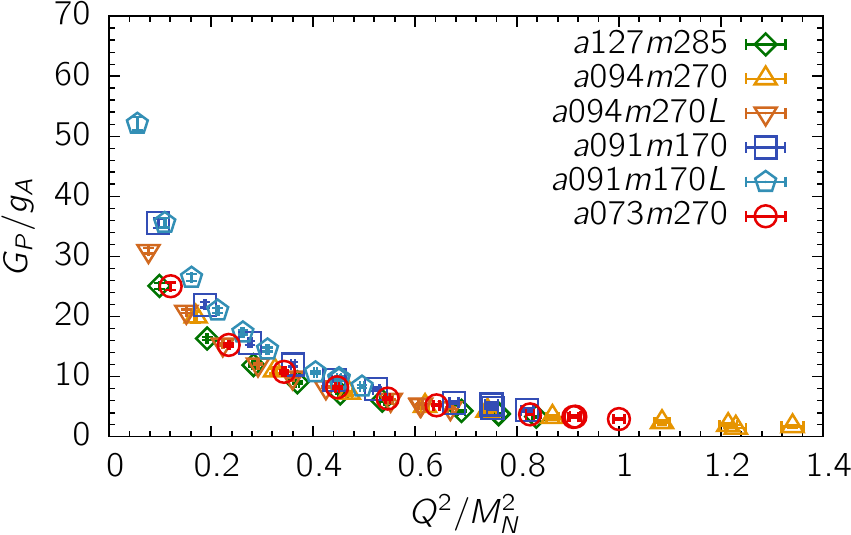}

  \includegraphics[width=0.325\textwidth]{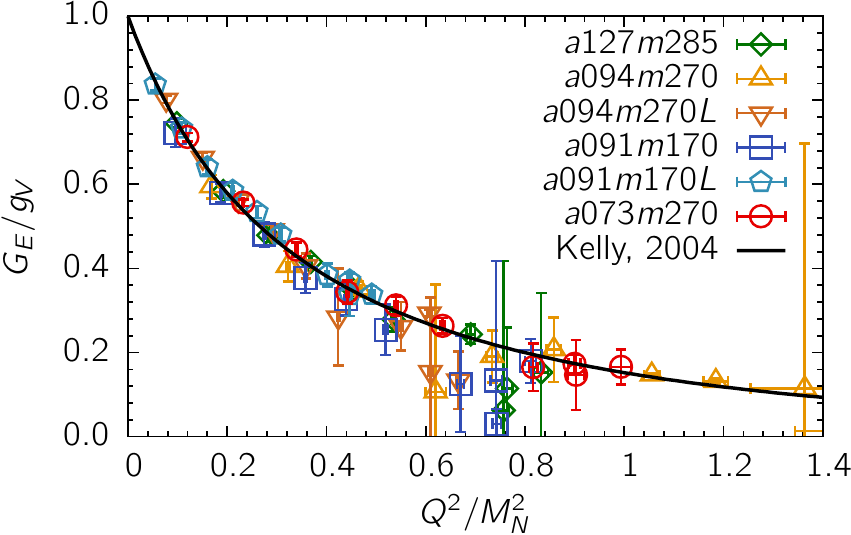}
  \includegraphics[width=0.325\textwidth]{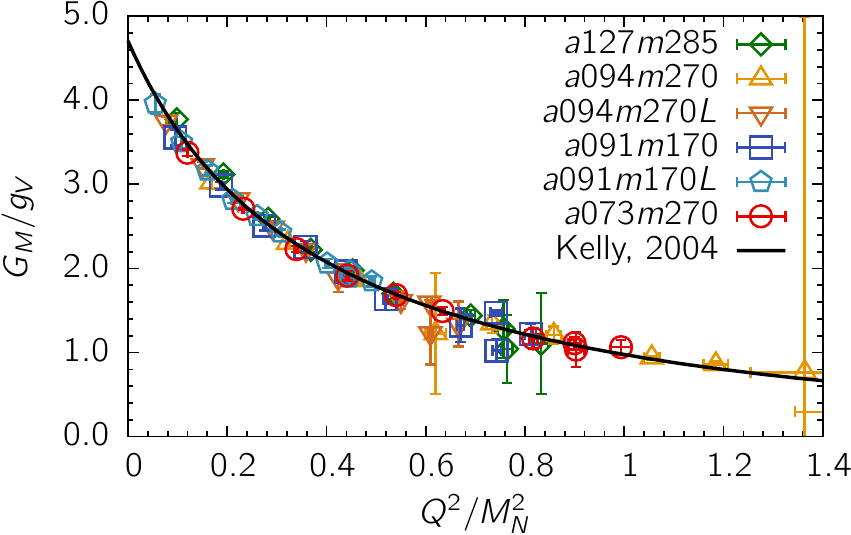}

  \includegraphics[width=0.325\textwidth]{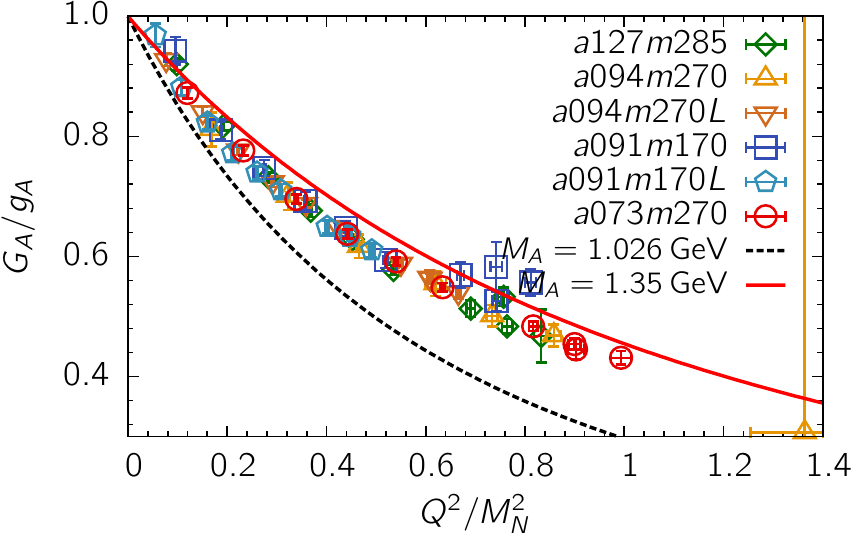}
  \includegraphics[width=0.325\textwidth]{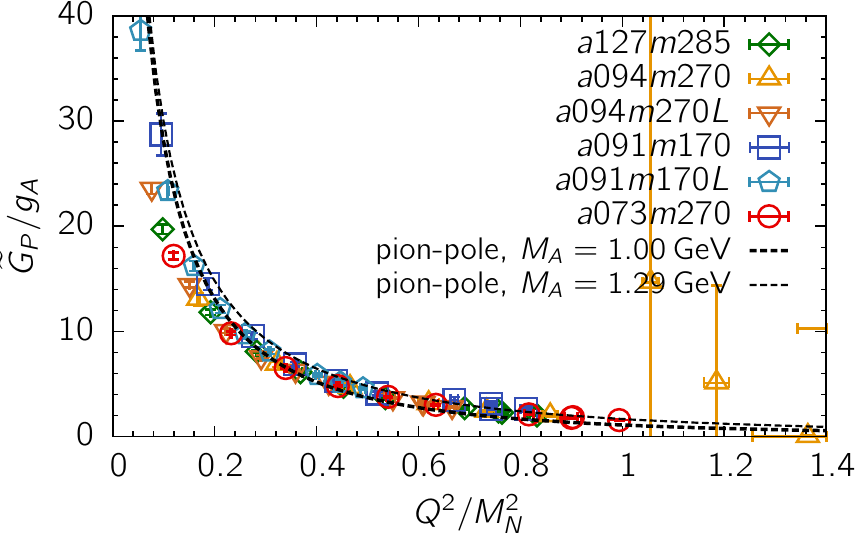}
  \includegraphics[width=0.325\textwidth]{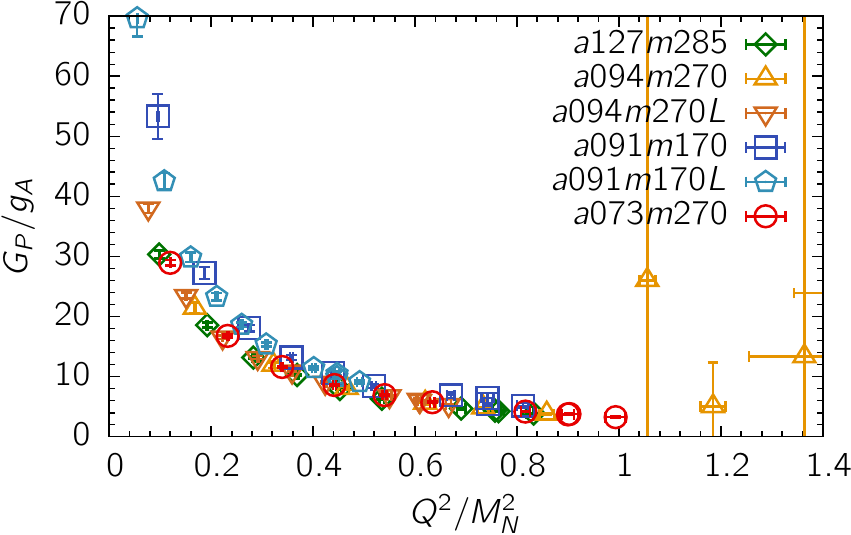}
  \vspace{-2mm}
  \caption{Renormalized isovector form factors ($G_E^{u-d}$,
    $G_M^{u-d}$, $G_A^{u-d}$, $\widetilde G_P^{u-d}$ and $G_P^{u-d}$)
    versus $Q^2/M_N^2$. Top (bottom) 5 panels show data with standard $\{4,3^\ast\}$ 
    (new) strategy. The value of $g_A$ is taken from $\{4,3^\ast\}$ fit. }
  \label{figs:clov-FF}
\end{figure}

We pointed out in Ref.~\cite{Jang:2019vkm} that the large violation of
the PCAC relation between axial and pseudoscalar form factors observed
in~\cite{Rajan:2017lxk} is due to lower energy $N\pi$ excited states
that are not exposed by the $\{4,3^\ast\}$ analysis.  Including them
addressed the PCAC relation \cite{Jang:2019vkm}. We, therefore explore
2 fit strategies here. The top 5 panels in Fig.~\ref{figs:clov-FF} show
renormalized $G_E^{u-d}$, $G_M^{u-d}$, axial ($G_A^{u-d}$), induced
pseudoscalar ($\widetilde G_P^{u-d}$) and pseudoscalar ($G_P^{u-d}$)
form factors analyzed using the standard $\{4,3^\ast\}$ strategy. The
bottom 5 panels are with (i) 2-state simultaneous fit to all $V_\mu$
channels for $G_E$ and $G_M$ with $E_1$ left free, and (ii) the ${
  S}_{A4}$ strategy defined in~\cite{Jang:2019vkm} for the axial
channels.

The $G_E$ and $G_M$ data show better collapse onto a single curve (indicating no significant  $a$, $M_\pi$, volume  dependence)
plotted versus $Q^2/M_N^2$, and the agreement with the Kelly curve is
better compared to the clover-on-HISQ data discussed in
Ref.~\cite{Jang:2019jkn}.  The main difference between the two
strategies is in the errors: the errors from the simultaneous fits are larger,
especially at the larger $Q^2$. \looseness-1

In the axial channels, data with ${S}_{A4}$ satisfies PCAC with most
of the change occuring in $\widetilde G_P^{u-d}$ and $G_P^{u-d}$ as
discussed in~\cite{Jang:2019vkm}. Note that data for 
$G_A^{u-d}$, $\widetilde G_P^{u-d}$ and $G_P^{u-d}$, shown in
Fig.~\ref{figs:clov-FF}, will move up or down depending on the value of
$g_A$, which, as shown in Tab.~\ref{tab:IVcharges}, has unresolved
systematics. Thus, resolving the ESC in $g_A$ is essential before 
comparing/using $G_A(Q^2)$ in phenomenology.

%%============================================================================

\begin{figure}[t]
  \center
  \vspace{-3mm}
  \includegraphics[width=0.46\textwidth]{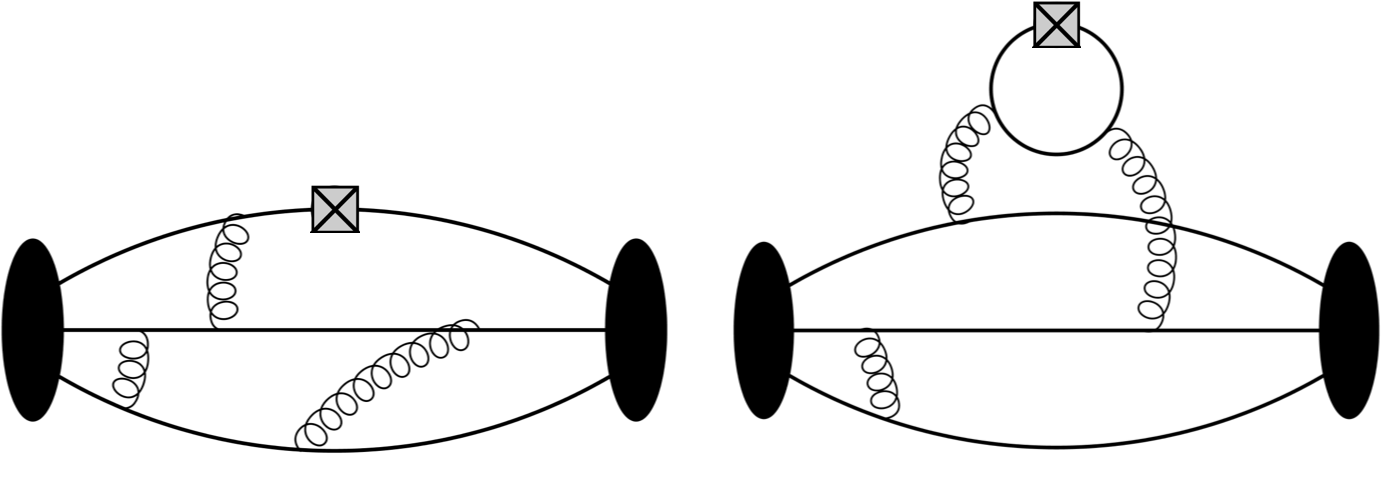}
  \hfill
  \includegraphics[width=0.23\textwidth]{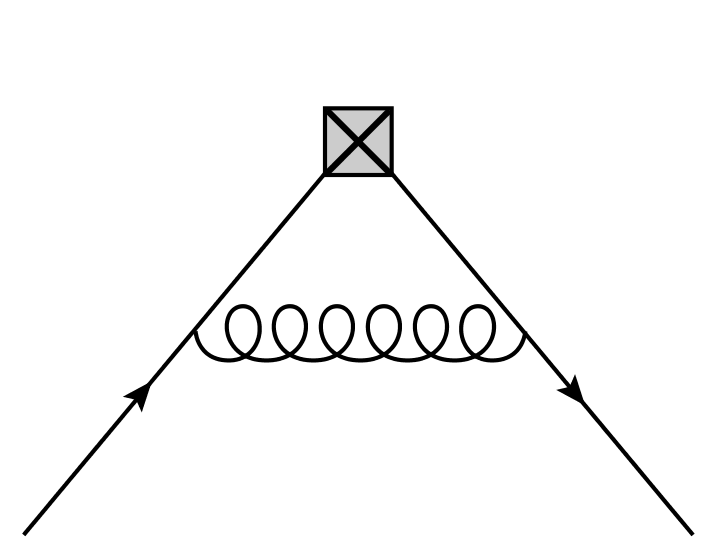}
  \includegraphics[width=0.23\textwidth]{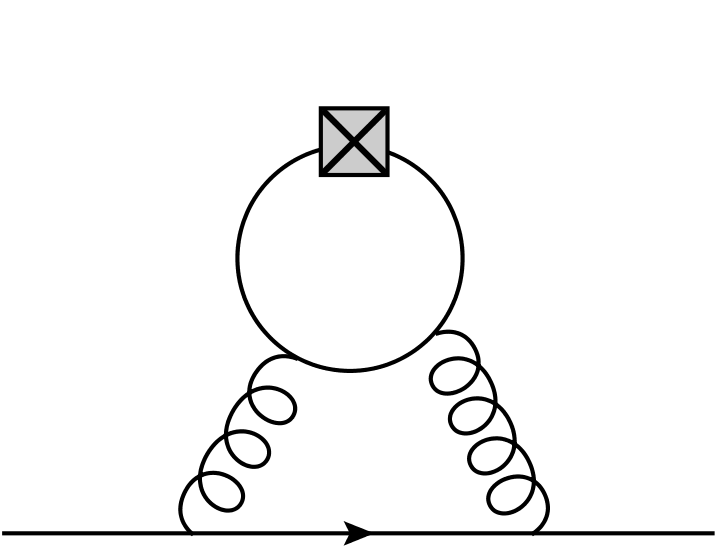}
  \vspace{-2mm}
  \caption{Connected and disconnected diagrams for (i) the 3-point
    functions that give nucleon
    charges (left 2 panels) and (ii) the renormalization of the flavor
    diagonal quark bilinear operators in 3-flavor theory (right 2).\looseness-1}
  \label{fig:disc}
\end{figure}

%%============================================================================
%%============================================================================
\section{Flavor diagonal charges on $2+1+1$-flavor HISQ lattices}
%%============================================================================
%%============================================================================

The flavor diagonal charges presented here are obtained using the same
ESC strategy as discussed in~\cite{Lin:2018obj,Gupta:2018lvp}.
Alternate analyses taking into account possible lower excited states
are in progress. The connected and disconnected contributions shown in
Fig.~\ref{fig:disc} are analyzed separately to construct the
renormalized charges $g_{A,S,T}^{f}|_R = Z^{ff^\prime}_{A,S,T}
(g_{A,S,T}^{f^\prime,\text{conn}} + g_{A,S,T}^{f^\prime,\text{disc}})$, where
$f,f^\prime$ are quark flavors. The connected contribution,
$g_{A,S,T}^{f,\text{conn}}$, are taken from Ref.~\cite{Gupta:2018qil}.
Here, we update $g_\Gamma^{f,\text{disc}}$ using the larger data set
shown in Table~\ref{tab:hisq}, and present new results on the
connected and disconnected contributions (right two panels in
Fig.~\ref{fig:disc}) for the renormalization matrix
$Z^{ff^\prime}_{A,S,T}$ in the 3-flavor theory using the RI-sMOM
scheme. The matching between the lattice RI-sMOM and continuum $\MSb$
schemes, and the running to 2~GeV are done using 2-loop perturbation
theory. Additionally, we give our first preliminary data for the
scalar charges.

%%============================================================================

\begin{figure}
  \center
  \vspace{-6mm}
  \includegraphics[width=0.263\textwidth]{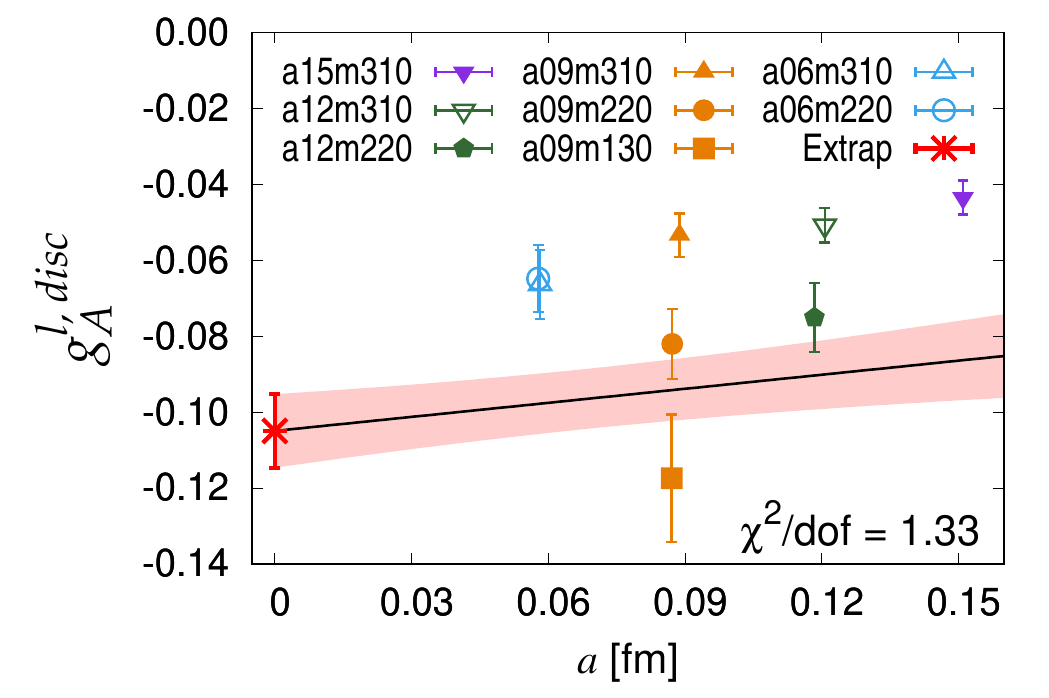}
  \includegraphics[width=0.227\textwidth, trim={13mm 0 0 0},clip]{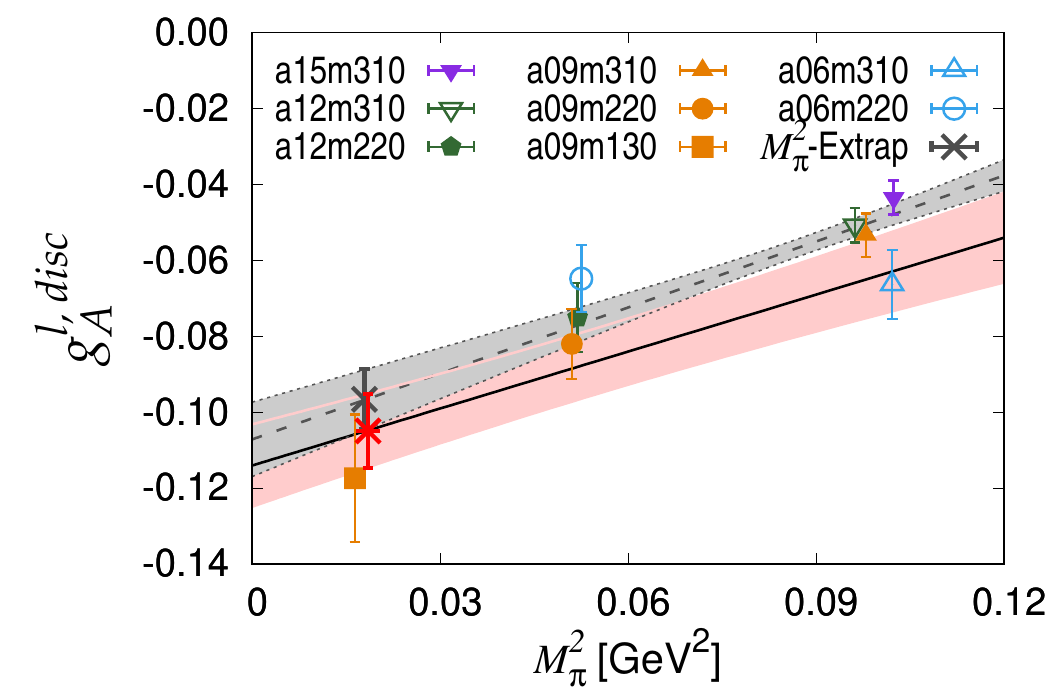}
  \includegraphics[width=0.263\textwidth]{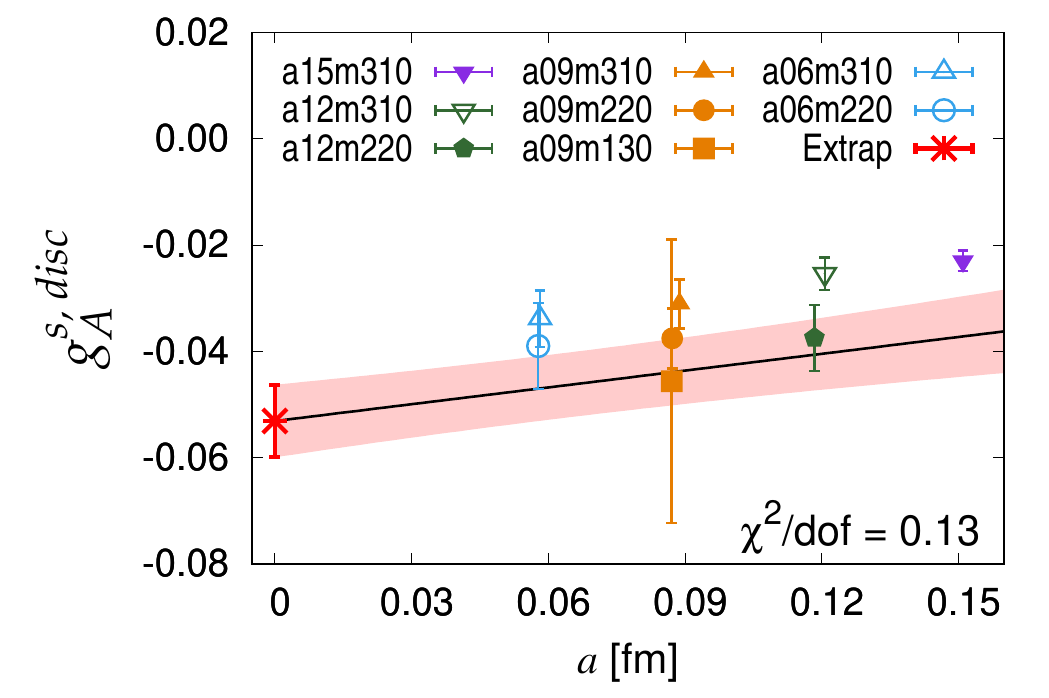}
  \includegraphics[width=0.227\textwidth, trim={13mm 0 0 0},clip]{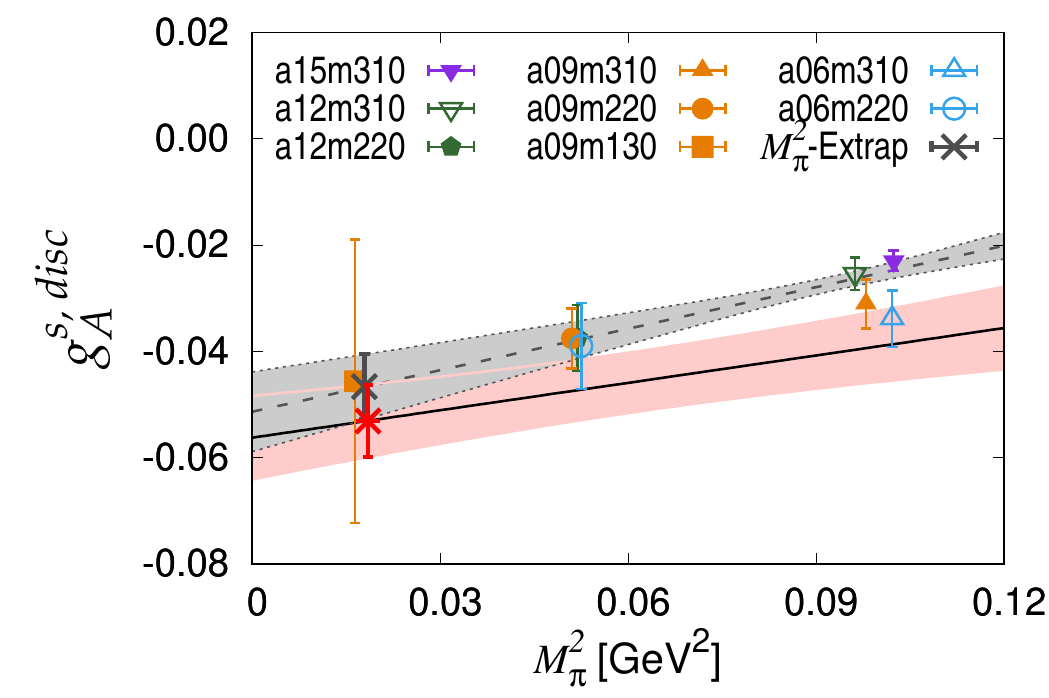}
  
  \includegraphics[width=0.26\textwidth]{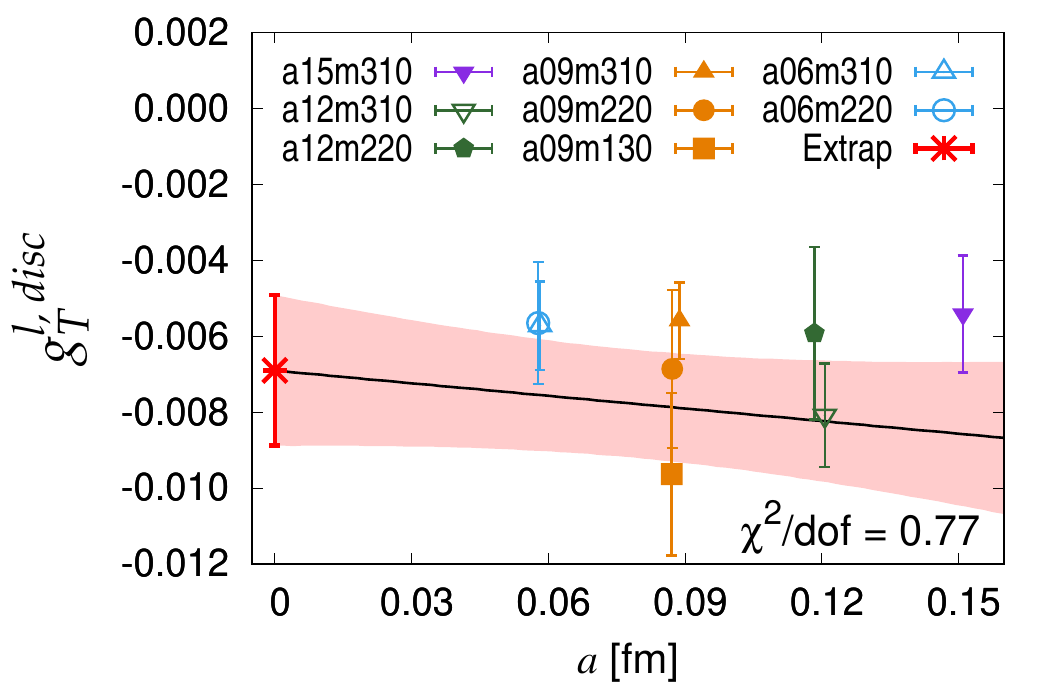}
  \includegraphics[width=0.23\textwidth, trim={12mm 0 0 0},clip]{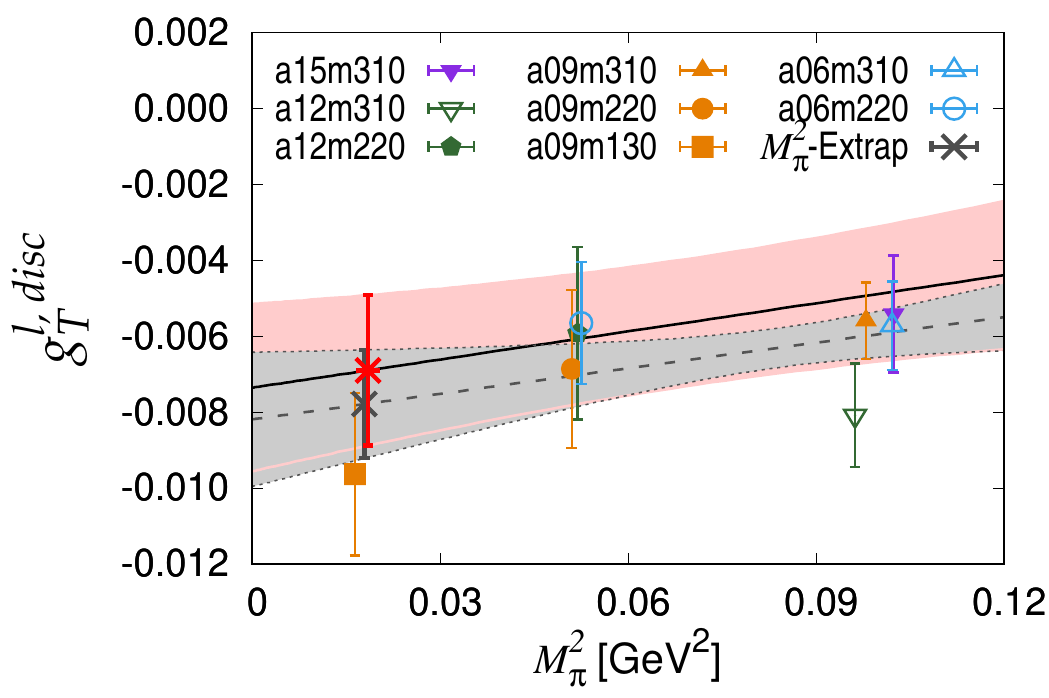}
  \includegraphics[width=0.26\textwidth]{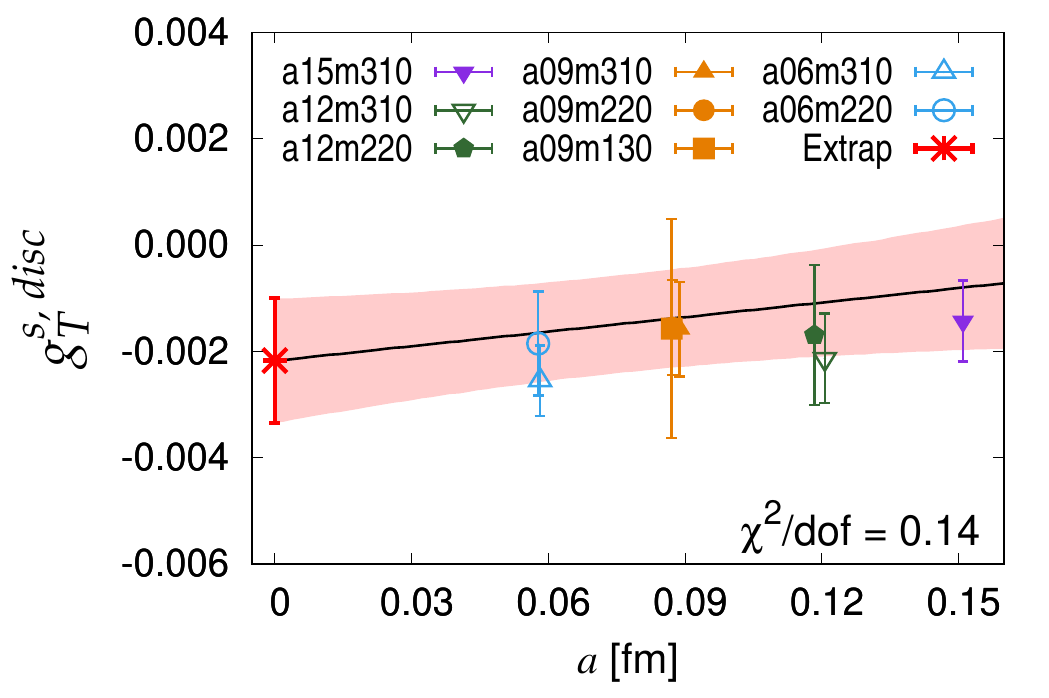}
  \includegraphics[width=0.23\textwidth, trim={12mm 0 0 0},clip]{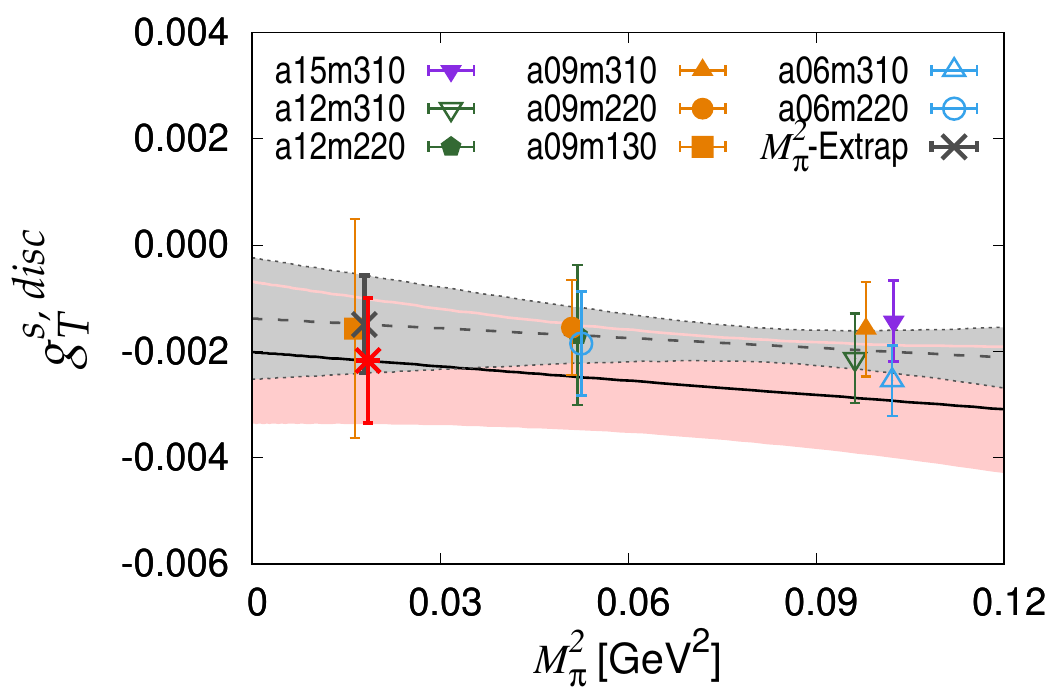}
  \vspace{-4mm}
  \caption{The extrapolation of the disconnected contributions of the
    renormalized (in $\MSb$ at $2$ GeV) flavor diagonal charges
    $g_A^{l(s),\text{disc}}|_R$ (top row) and
    $g_T^{l(s),\text{disc}}|_R$ (bottom row) using the
    chiral-continuum fit ansatz $g(a,M_\pi)=c_1+c_2a +c_3
    M_\pi^2$. The parameters for the eight clover-on-HISQ ensembles 
    are given in Table~\ref{tab:hisq}. }
  \label{figs:disc_extrap}    
\end{figure}

%
The new data for $Z$ factors in Table \ref{tab:Z} show that the difference between the
isovector ($u-d$) and isoscalar ($u+d$) renormalization constants for
the axial and tensor operators is small for all 4 values of $a$. This
validates the approximation
$g_{A,T}^{u+d}|_R={Z_{A,T}^{u+d,u+d}}g_{A,T}^{u+d} +
Z_{A,T}^{u+d,s}g_{A,T}^{s}\approx Z_{A,T}^{u-d}g_{A,T}^{u+d}$ made in
Refs.~\cite{Lin:2018obj,Gupta:2018lvp} for our
clover-on-HISQ calculations.  Also, the off-diagonal mixing
$Z_{A,T}^{u+d,s}g_{A,T}^{s}$ is tiny since $Z_{A,T}^{u+d,s}\lesssim
0.1$ and the purely disconnected contributions are even smaller:
$g_{A}^{s}\approx 0.05$ and $g_{T}^{s}\approx 0.002$.
With only 8 data points, the chiral-continuum extrapolations of the disconnected contributions
$g_{A,T}^{l,s,\text{disc}}|_R$ are carried out using the simple ansatz 
$g(a,M_\pi)=c_1+c_2a +c_3 M_\pi^2$ and the data and fits are shown in 
Fig~\ref{figs:disc_extrap}. Combining the disconnected contributions
with the connected contributions $g_{A,T}^{l,\text{conn}}|_R$ presented in Ref.~\cite{Gupta:2018qil},
our preliminary updated flavor diagonal charges are
\begin{align}
  g_A^{u} |_R&= 0.790(23)(30)& g_A^{d}|_R&=-0.425(15)(30)& g_A^{s}|_R&=-0.053(7)\\
  g_T^{u} |_R&= 0.783(27)(10)& g_T^{d}|_R&= -0.205(10)(10)& g_T^{s}|_R&= -0.0022(12) \,,
\end{align}
where the second is a systematic error assigned to the chiral-continuum extrapolation~\cite{Gupta:2018qil}.
\begin{table}
  \vspace{-7mm}
  \center
  \resizebox{.34\textwidth}{!}{
    \begin{tabular}{c|cc}
      a & $Z_A^{u-d}/{Z_V^{u-d}}$ & ${Z_A^{u+d,u+d}}/{Z_V^{u-d}}$\\\hline
      0.15 & 1.080(13) & 1.0856(11)\\
      0.12 & 1.061(11) & 1.073(11)\\
      0.09 & 1.0380(40) & 1.0484(32)\\
      0.06 & 1.0227(19) & 1.0383(28)
    \end{tabular}}
    \resizebox{.34\textwidth}{!}{
    \begin{tabular}{c|cc}
      a & $Z_T^{u-d}/{Z_V^{u-d}}$ & ${Z_T^{u+d,u+d}}/{Z_V^{u-d}}$\\\hline
      0.15 & 1.032(18) & 1.031(19)\\
      0.12 & 1.0538(76) & 1.0541(68)\\
      0.09 & 1.0795(34) & 1.0796(34)\\
      0.06 & 1.0956(70) & 1.0959(70)\\
    \end{tabular}}
    \caption{Renormalization factors $Z_A$ and $Z_T$ for $u-d$
      (isovector) and $u+d$ (isoscalar) operators in $\MSb$ scheme at
      $2$~GeV on HISQ lattices. Renormalizing using ratios with
      $Z_V^{u-d}$ is intended to cancel some of the statistical and
      systematic uncertainties as discussed in
      Ref.~\cite{Gupta:2018qil}.  Errors quoted are the larger of the
      two: half the difference between RI-MOM and RI-sMOM results or
      the largest statistical error. }
    \label{tab:Z}
    \vspace{-2mm}
\end{table}
%%============================================================================

There remain issues regarding the systematics in the
calculation of the matrix element of the scalar operator that are
still being investigated: the values for the the renormalization constants,
$Z_S$, show significant differences between the RI-MOM and RI-sMOM
schemes. For example, there are $5\sim30\%$ differences in
$Z_S^{u-d}$, and $5\sim 10\%$ differences in $Z_S^{u+d,u+d}$ with the
differences increasing as the lattice spacing becomes larger. For the
mixing matrix element $Z_S^{s,u+d}$, the RI-MOM scheme gives $-0.04
\sim -0.1$, which is much larger than the RI-sMOM scheme result of
$-0.003\sim -0.02$. In fact, in the calculation of strangeness,
$g_S^{s}|_R = {Z_{S}^{s,s}}g_{S}^{s} + Z_{S}^{s,u+d}g_{S}^{u+d}$, the
larger value of mixing $Z_{S}^{s,u+d}$ in RI-MOM scheme gives a
negative value for $g_S^{s}|_R$! For the time being, we
use the RI-sMOM scheme, in which case the corresponding mixing
term $Z_{S}^{s,u+d}g_{S}^{u+d}$ gives about $6\sim20\%$ correction to
the diagonal term ${Z_{S}^{s,s}}g_{S}^{s}$.

\begin{figure}
  \center
  \includegraphics[width=0.37\textwidth]{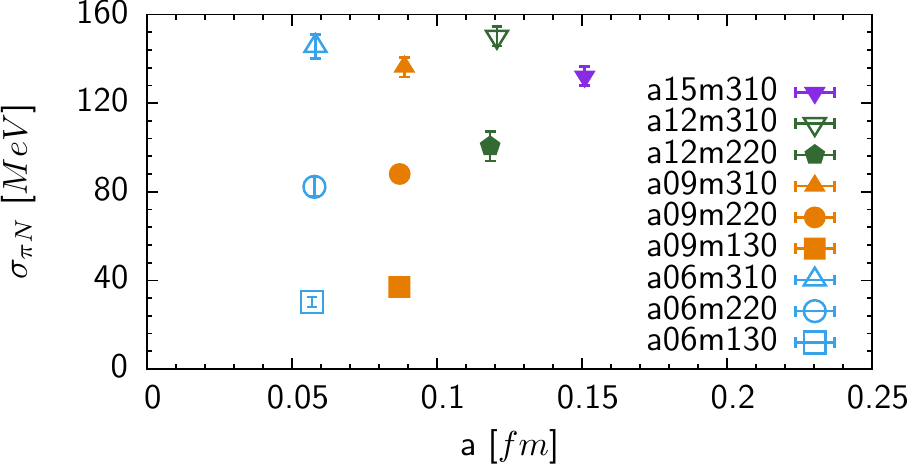}
  \includegraphics[width=0.37\textwidth]{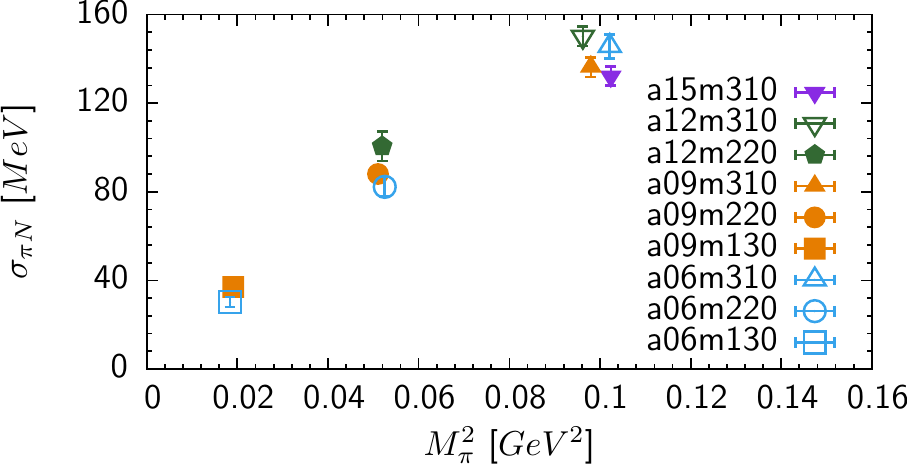}
  \includegraphics[width=0.37\textwidth]{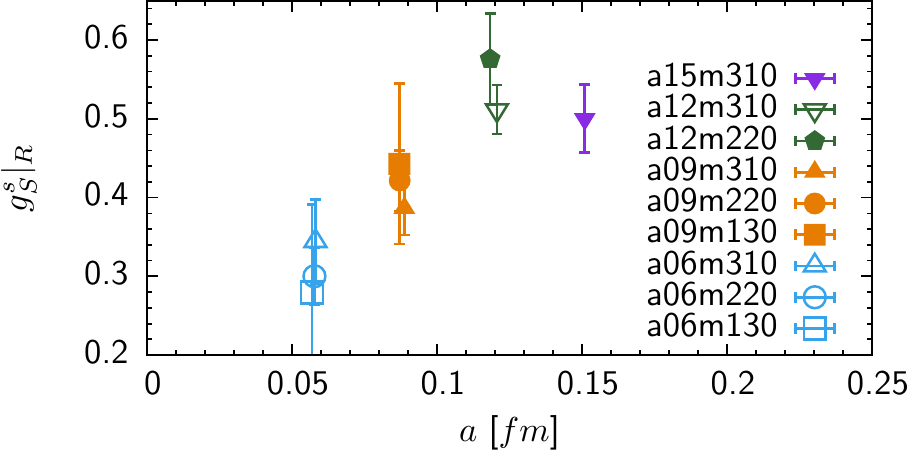}
  \includegraphics[width=0.37\textwidth]{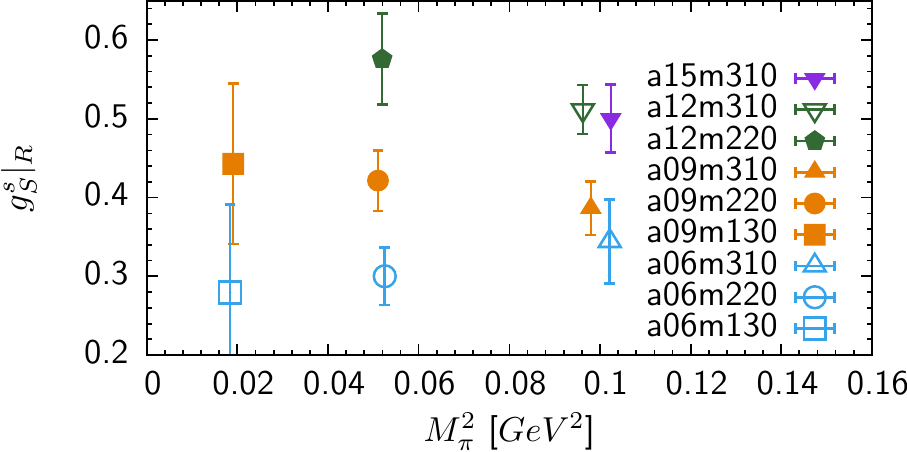}
  \vspace{-2mm} 
  \caption{(Top) The nucleon sigma term $\sigma_{N\pi}$ plotted versus $a$ and
    $M_\pi^2$. (Bottom) The nucleon strangeness $g_s^{s}|_{R}$ renormalized in $\MSb$ scheme at $2$ GeV versus
    $a$ and $M_\pi^2$. }
  \label{figs:S}
\end{figure}

The renormalized strangeness $g_S^{s}|_R$, from the clover-on-HISQ
calculation, is plotted versus $a$ and $M_\pi^2$ in Fig.~\ref{figs:S},
along with the nucleon sigma term $\sigma_{\pi N}=m_lg_S^{u+d}$ that
is independent of the renormalization scheme. We have used $a
m_l=\frac{1}{2}(\kappa^{-1}-\kappa_\text{crit}^{-1})$ for the
definition of the bare quark mass and
$g_S^{u+d}=g_S^{u+d,\text{conn}}+g_S^{u+d,\text{disc}}$ for the unrenormalized
isoscalar scalar charge. The data show a significant $a$ dependence in
$g_S^{s}|_R$ while the large linear dependence of $\sigma_{\pi N}$ on
$M_\pi^2$ comes from the quark mass in the definition of $\sigma_{\pi
  N}$. The dependence of $g_S^{u+d}|_R$ on $a$ and of $g_S^{s}|_R$ on
$M_\pi^2$ is not clear.\looseness-1

%%============================================================================

%%============================================================================
%%============================================================================
%%  \section{Conclusion and future works}
%%============================================================================
%%============================================================================

{\bf Conclusions:} We have presented the status of ongoing
calculations of nucleon matrix elements and are performing a more
detailed analysis of the excited state contamination in the extraction
of {\bf all} nucleon matrix elements.  The analysis of flavor diagonal
scalar charges, $g^{u,d,s}_{S}$ is new, however, a complete
understanding of all the systematics is still under investigation.

{\bf Acknowledgments:} 
We thank the MILC collaboration for sharing their $2+1+1$-flavor HISQ
ensembles. 
Simulations were carried out on computer facilities at (i) the
National Energy Research Scientific Computing Center, a DOE Office of
Science User Facility supported under Contract No. DE-AC02-05CH11231;
and, (ii) the Oak Ridge Leadership Computing Facility supported by the
Office of Science of the DOE under Contract No. DE-AC05-00OR22725;
(iii) the USQCD Collaboration, which are funded by the Office of
Science of the U.S. Department of Energy, and (iv) Institutional
Computing at Los Alamos National Laboratory. T.~Bhattacharya and
R.~Gupta were partly supported by the U.S. Department of Energy,
Office of Science, Office of High Energy Physics under Contract
No. DE-AC52-06NA25396. S.~Park, T.~Bhattacharya, R.~Gupta, Y.-C.~Jang
and B.~Yoon were partly supported by the LANL LDRD program.

%% \end{thebibliography}
%%%%%%%%%%%%%%%%%%%%%%%%%%%%%%%%%%%%%%%%%%%%%%%%%%%%%%%%%%%%%%%%%%%%%%%
\bibliographystyle{JHEP}
\bibliography{ref} %%% ref.bib file

\providecommand{\href}[2]{#2}\begingroup\raggedright\begin{thebibliography}{1}

\bibitem{Jang:2019vkm}
Y.-C. Jang, R.~Gupta, B.~Yoon, and T.~Bhattacharya
  \href{http://xxx.lanl.gov/abs/1905.06470}{{\tt 1905.06470}}.

\bibitem{Bazavov:2012xda}
{\bf MILC} Collaboration, A.~Bazavov {\em et~al.} {\em Phys. Rev.} {\bf D87}
  (2013), no.~5 054505, [\href{http://xxx.lanl.gov/abs/1212.4768}{{\tt
  1212.4768}}].

\bibitem{Lin:2018obj}
H.-W. Lin, R.~Gupta, B.~Yoon, Y.-C. Jang, and T.~Bhattacharya {\em Phys. Rev.}
  {\bf D98} (2018), no.~9 094512,
  [\href{http://xxx.lanl.gov/abs/1806.10604}{{\tt 1806.10604}}].

\bibitem{Gupta:2018lvp}
R.~Gupta, B.~Yoon, T.~Bhattacharya, V.~Cirigliano, Y.-C. Jang, and H.-W. Lin
  {\em Phys. Rev.} {\bf D98} (2018), no.~9 091501,
  [\href{http://xxx.lanl.gov/abs/1808.07597}{{\tt 1808.07597}}].

\bibitem{Yoon:2016jzj}
B.~Yoon {\em et~al.} {\em Phys. Rev.} {\bf D95} (2017), no.~7 074508,
  [\href{http://xxx.lanl.gov/abs/1611.07452}{{\tt 1611.07452}}].

\bibitem{Gupta:2018qil}
R.~Gupta, Y.-C. Jang, B.~Yoon, H.-W. Lin, V.~Cirigliano, and T.~Bhattacharya
  {\em Phys. Rev.} {\bf D98} (2018) 034503,
  [\href{http://xxx.lanl.gov/abs/1806.09006}{{\tt 1806.09006}}].

\bibitem{Rajan:2017lxk}
R.~Gupta, Y.-C. Jang, H.-W. Lin, B.~Yoon, and T.~Bhattacharya {\em Phys. Rev.}
  {\bf D96} (2017), no.~11 114503,
  [\href{http://xxx.lanl.gov/abs/1705.06834}{{\tt 1705.06834}}].

\bibitem{Jang:2019jkn}
Y.-C. Jang, R.~Gupta, H.-W. Lin, B.~Yoon, and T.~Bhattacharya {\em Phys. Rev.}
  {\bf D101} (2020) 014507, [\href{http://xxx.lanl.gov/abs/1906.07217}{{\tt
  1906.07217}}].

\end{thebibliography}\endgroup
%%%%%%%%%%%%%%%%%%%%%%%%%%%%%%%%%%%%%%%%%%%%%%%%%%%%%%%%%%%%%%%%%%%%%%%

\end{document}